\documentclass[aps,prl,twocolumn,groupedaddress,preprintnumbers,showpacs,amsmath,amssymb]{revtex4}

\usepackage{bm}

\begin{document}

\preprint{Submitted to Phys. Rev. Lett.}

\title{Pressure-driven Instabilities in Cylindrical Geometry:
\\ A New General Criterion}


\author{Pierre-Yves Longaretti}
\email{pyl@obs.ujf-grenoble.fr}
\homepage{http://www-laog.obs.ujf-grenoble.fr/~pyl/}
\affiliation{Laboratoire d'Astrophysique de Grenoble, BP 53X,
F-38041, Grenoble, France}

%\thanks

\date{03/20/03}

\begin{abstract}

A new criterion for pressure-driven interchange instabilities in
cylindrical geometry is derived, based on an alternate use of the
Energy Principle. This criterion is inequivalent to Suydam's
criterion and does not contain the magnetic shear. In fact, it is
shown that Suydam's criterion relates to the instability of the
slow magnetosonic branch, while the present criterion relates to
the Alfv\'enic one, which is the most dangerous of the two. These
findings imply that pressure-driven modes often exist even if
Suydam's criterion is satisfied by a large margin.

\end{abstract}


%\keywords{}

\maketitle

MHD instabilities are usually divided into pressure-driven and
current-driven, depending on whether the destabilizing agent
results from the interplay between the gas pressure gradient and
the field-line curvature, or from the component of the current
parallel to the magnetic field. The stability in cylindrical
geometry with respect to both types of processes is an old and
well-trodden subject. It was actively studied in the 60s and 70s,
both in its own sake and as an idealization of more complex
geometries for fusion devices. Important general results have been
established relatively early on, and are now part of the standard
lore on the subject \cite{Freid87}; among these Suydam's criterion
\cite{suyd58} for interchange pressure-driven modes in an
arbitrary screw pinch is directly related to the present work.
Later developments have mostly been concerned with the stability
properties of particular configurations such as tokamaks or
reversed-field pinches.

In addition to fusion devices, astrophysical jets constitute
another class of physical systems for which the existence of
purely MHD unstable modes in cylindrical geometry is extremely
relevant. However, there are only a handful of papers on
current-driven instabilities in jets (\cite{AC92,Ap96,ALB}, and
references therein), and only a couple of papers on
pressure-driven instabilities \cite{Beg98,KLP00}. This
surprisingly small level of activity reflects the fact that the
Kelvin-Helmholtz instability has long been thought to be the most
important and dangerous for jet survival, a conjecture somewhat
belied by recent numerical results \cite{RJF00}.

Suydam's criterion stresses that the magnetic shear is an
important stabilizing factor in cylindrical MHD columns. Indeed,
although this criterion is only a necessary condition for
stability, it is well-known that the magnetic shear plays an
important role in the ability of nuclear fusion devices to survive
the catastrophic disruption usually produced by MHD instabilities.
However, it has also long been noticed that pressure-driven
unstable modes are easily found in MHD equilibria, with
growth-rates nearly independent of their wavenumber, independently
of whether Suydam's criterion is satisfied or not (see, e.g.,
Ref.~\cite{Mer89}).

One of the objectives of this Letter is to provide a theoretical
explanation for this fact, and to show that the role of
pressure-driven interchange modes has probably been underestimated
in fusion physics. This is done through the derivation of a new
general criterion for the stability of pressure-driven modes, and
by a discussion of the physical meaning of this new criterion as
well as Suydam's. The criterion is obtained from an alternate use
of the MHD Energy Principle \cite{BFKK,Lav,Freid87}.

For static equilibria, the MHD perturbation equations can be
written as

\begin{equation}\label{mvt}
  -\rho_o \omega^2{\bm\xi}=-{\bm \nabla} \delta P_\ast+ \delta{\bm T}
  \equiv {\bm F}({\bm \xi}) ,
\end{equation}

\noindent where $\delta{\bm T}$ is the perturbation of the
magnetic tension, and $\delta P_\ast$ the total pressure
perturbation, and $\bm\xi$ the fluid element displacement from
equilibrium (a Fourier transform in time has been performed).
Specializing to cylindrical equilibria, it is convenient to
project this perturbation equation in the three directions ${\bm
e}_r$, ${\bm e}_\parallel$, ${\bm e}_\perp$ (i.e., the radial
direction, the direction parallel to the unperturbed magnetic
field, and the remaining one required to define an orthonormal
frame); furthermore the displacement can be chosen as a Fourier
mode:

\begin{equation}\label{equ:xif}
{\bm\xi}={\bm\xi}(r)\exp\left[i(\omega t-m\theta-k_z z)\right].
\end{equation}

This choice reflects the fact that the equilibrium quantities (the
magnetic field line helical structure, the magnetic field
components and the gas pressure) depend only on radius. The
equations for the parallel and perpendicular displacement
$\xi_\parallel$ and $\xi_\perp$ can be solved exactly as a
function of $\omega^2$ and $\xi_r$ \cite{HL58,Freid87,Long03}:

\begin{equation}\label{xipar}
  \xi_{\parallel,\perp}=\xi_{\parallel,\perp}(\omega^2,\xi_r),
\end{equation}

\noindent leading to an exact second order equation for the radial
displacement \cite{HL58,Freid87,Long03}:

\begin{equation}
   \label{xir}
-\rho\omega^2\xi=
\frac{d}{dr}\left[A\frac{d}{dr}(r\xi_r)\right]+C^*r\xi_r\equiv
F_r(\xi_r).
\end{equation}

\noindent In this expression, $A$ and $C^*$ depend in a
complicated way on the equilibrium quantities and on the mode
frequency $\omega$.

Let us define

\begin{equation}\label{dWr}
\delta W_r(\xi_r^*, \xi_r)=-\frac{1}{2}\int\xi_r^*
F_r({\bm\xi})d^3 r,
\end{equation}

\noindent and

\begin{equation}\label{dKr}
\delta K_r(\xi_r^*, \xi_r)=\frac{1}{2}\int\rho_o|\xi_r|^2 d^3 r.
\end{equation}

\noindent where, for any choice of $\xi_r$, the remaining
displacement components $\xi_\parallel$ and $\xi_\perp$ are
specified by Eqs.~(\ref{xipar}), and $\omega^2=\omega^2[\xi_r]$ is
chosen to be the formal solution of the implicit equation

\begin{equation}\label{omnegr}
\omega^2={{\delta W_r}\over {\delta K_r}}.
\end{equation}

 It can be shown from the standard form of the Energy Principle that,
if there exists a total displacement such that $\delta W_r < 0$,
then the system is unstable; furthermore, $\omega^2$, as given by
Eq.~(\ref{omnegr}), also satisfies $\omega^2=\delta W/\delta K$,
where $\delta W$ and $\delta K$ are the usual potential and
kinetic energy of the standard Energy Principle \cite{Long03}.

Obviously, this reexpression of the Energy Principle is useful
only if Eq.~(\ref{omnegr}) can be solved for $\omega^2$, which is
in general a rather daunting task. Fortunately, an interesting and
useful simplification of $F_r$ is suggested by the ballooning
asymptotic ordering. In particular, let us look for trial
displacements where $m,k_z$ satisfy $k_\perp L_o \gg 1$ while
$k_\parallel L_o \ll 1$ ($L_o$ is the characteristic scale of
equilibrium gradients). In other words, one considers
displacements with large wavenumbers, except in the direction of
the unperturbed magnetic field. Because of the magnetic shear,
this imposes that the displacement must be nonvanishing only in
the close vicinity to the magnetic resonance associated to $m,k$,
and implicitly defined by $k_\parallel=0$. Furthermore, let us
make the assumption that one can find a trial radial displacement
$\xi_r$ such that the frequency $\omega^2$ which is the solution
of Eq.~(\ref{omnegr}) obeys the ballooning ordering, i.e.
$\omega_f^2\gg |\omega^2|\ (\sim v_A^2/L_o^2, c_s^2/L_o^2)\ \gg
v_A^2 k_\parallel^2, c_s^2 k_\parallel^2$, where $\omega_f^2\simeq
(v_A^2+c_s^2)k_\perp^2$ is the fast magnetosonic frequency. This
assumption is easily checked a posteriori.

With these approximations, $k_\parallel$ can be expanded to first
order with respect to the distance to the magnetic resonance
$r_o$, to read

\begin{equation}\label{ko'}
  k_\parallel= -k_\perp\frac{B_\theta B_z}{B_o^2}\left(\frac{r-r_o}{r}\right)\frac{d\ln |h|}
  {d\ln r},
\end{equation}

\noindent where $h\equiv rB_z/B_\theta$ is the pitch length (its
logarithmic derivative is the magnetic shear), and $B_\theta, B_z,
B_o$ are the equilibrium azimuthal, vertical, and total magnetic
field, respectively. The width of the region where the radial
displacement is nonvanishing can be quantified from the constraint
that $k_\parallel L_o$ is small enough (see below) with the help
of this expression. For large $k_\perp L_o$, this width is narrow
enough so that one can neglect the variations with radius of all
quantities (except $k_\parallel$) in the corresponding region.

From Eq.~(\ref{ko'}) (and the approximations just described), the
quantity $C^*$ reduces to \cite{Long03}

\begin{equation}\label{equ:C'}
  C^*=-\frac{2B_{\theta}^2}{r^3}\left(\frac{rP_o'}{B_o^2}\right)-
  \frac{4\beta B_{\theta}^4}{(1+\beta) r^3
  B_o^2}\equiv\frac{\omega_c^2}{r}.
\end{equation}

In this expression, the system of units is chosen such that
$\mu_o=1$, and the plasma $\beta$ parameter is defined as
$\beta\equiv c_s^2/v_A^2$, i.e. as the square of the ratio of the
adiabatic sound speed (for an adiabatic equation of state) to the
Alfv\'en speed; this definition differs from the standard one by a
factor $\gamma/2$ ($\gamma$ is the adiabatic index).

With these simplifications, Eq.~(\ref{omnegr}) now reads

\begin{equation}\label{critint}
\omega^2\int \rho_o|\xi_r|^2=\int
Ar\left|{{d\xi_r}\over{dr}}\right|^2
  d r +\omega_c^2\int \rho_o|\xi_r|^2 dr,
\end{equation}

\noindent where, to leading order, $A$ is given by

\begin{equation}\label{A'}
A=-{\rho_o\over r}{\omega^2\over k_\perp^2}.
\end{equation}

Solving for $\omega^2$ yields:

\begin{equation}\label{omcrit}
\omega^2=\omega_c^2 {{\int_{r_o-\delta}^{r_o+\delta} |\xi_r|^2
  dr}\over{\int_{r_o-\delta}^{r_o+\delta} (|\xi_r|^2 +
  {1}/{{k_\perp}^2}|{d\xi_r}/{dr}|^2) dr}},
\end{equation}

\noindent where $\omega_c^2$ is defined in Eq.~(\ref{equ:C'}). The
quantity $\delta$ appearing in Eq.~(\ref{omcrit}) measures the
width of the region where the radial displacement is nonvanishing.
It must be chosen such that the constraint

\begin{equation}\label{delta}
v_A^2 k_\parallel^2, c_s^2 k_\parallel^2 \lesssim |\omega^2|
\end{equation}

\noindent is satisfied, with the help of Eq.~(\ref{ko'}) (if the
displacement is non zero in regions where $v_A^2 k_\parallel^2,
c_s^2 k_\parallel^2 \gtrsim |\omega^2|$, it produces positive
contributions to the potential energy $\delta W_r$ \cite{Long03}).
It is apparent that any displacement satisfying this constraint
makes $\omega^2 < 0$ provided that

\begin{equation}\label{crit}
  \omega_c^2=
  v_A^2\left(-2\beta\kappa_\rho\kappa_c+\frac{4\beta\kappa_c^2}{1+\beta}\right) <
  0.
\end{equation}

\noindent In Eq.~(\ref{crit}), $\kappa_c\equiv {\bm e}_r. [({\bm
e}_\parallel.{\bm\nabla}){\bm e}_\parallel]=-B_\theta^2/(B_o^2 r)$
is an algebraic measure of the field line inverse radius of
curvature, while $\kappa_\rho=(d\rho_o/dr)/\rho_o$ is a measure of
the pressure gradient scale (an adiabatic equation of state is
assumed). This reexpression of $\omega_c^2$ allows us to recognise
the first term as the usual destabilizing term due to the gas
pressure and the field line curvature, while the second
(stabilizing) term arises from the plasma compression. As a matter
of fact, this criterion generalizes a result due to Kadomtsev for
the saussage ($m=0$) mode in a Z-pinch \cite{KAD}.

The reader can easily convince himself that the condition required
earlier on $\omega^2$ is satisfied for reasonable choices of
$\xi_r$ if $\omega_c^2\neq 0$, thereby insuring the
self-consistency of the result. In the whole procedure, and in
contrast to common usage, neither the incompressibility condition
nor marginal stability has been assumed (actually, the two
conditions are known to be tied to one another). It is precisely
because of this feature that it was possible to derive the
criterion. Note that the usual reasoning according to which the
most dangerous modes are incompressible is valid only if
incompressible modes with finite growth rates can be found, which
is not the case for pressure-driven modes in MHD static
equilibria, as just pointed out.

Within the ballooning asymptotic ordering, there is another
well-known limiting case of direct interest to the purpose of this
Letter, that make Eq.~(\ref{xir}) analytically tractable: namely,
$\omega_f^2 \gg v_A^2/L_o^2, c_s^2/L_o^2 \gg v_A^2 k_\parallel^2,
c_s^2 k_\parallel^2\gg |\omega^2|\approx 0$. This limit allows us
to recover Suydam criterion. Indeed, in this limit, the fast
variation of $k_\parallel$ with radius dominates over all others,
and after some algebra, Eq.~(\ref{xir}) reduces to

\begin{equation}\label{equ:suyd}
\frac{d}{dr} \left(\frac{rB_o^2 k_\parallel^2}
{k^2}\frac{d\xi_r}{dr}\right) -\frac{2k_z^2 P_o'}{k^2}\xi_r=0,
\end{equation}

\noindent where $k\equiv (k_z^2+m^2/r^2)^{1/2}$. With the help of
Eq.~(\ref{ko'}), this equation gives the behavior of $\xi$ as a
function of radius as $\xi\propto [(r-r_o)/r_o]^p$; $p$ is real
when

\begin{equation}\label{crit:suyd}
  rB_z^2\left(\frac{h'}{h}\right)^2 + 8 P_o' > 0,
\end{equation}

\noindent $p$ is complex otherwise. In this last case, it is
possible to show, with the help of the Energy Principle, that the
fluid is unstable (see, e.g., Ref.~\cite{Freid87} for details).
Therefore, Eq.~(\ref{crit:suyd}) (Suydam's criterion
\cite{suyd58}) constitutes a necessary condition for stability.

This stresses the fact that both the new instability criterion
Eq.~(\ref{crit}), and Suydam's necessary stability criterion
Eq.~(\ref{crit:suyd}) follow from Eq.~(\ref{xir}), but in
different limits, and are not mutually exclusive. A deeper insight
into the relation between these two criteria is obtained by
considering the low magnetic shear limit. In this case, a WKB
approximation can be made on $\xi_r$ with radial wavenumber $k_r$
satisfying $|k_\perp r| \gg |k_r r| \gg 1$, so that the term
involving the derivative of $\xi_r$ can be neglected in
Eq.~(\ref{xir}). Assuming $\omega_f^2 \gg |\omega^2|, v_A^2
k_\parallel^2, c_s^2 k_\parallel^2$, Eq.~(\ref{xir}) reduces to
the following dispersion relation \cite{KLP00,Long03}
\footnote{There is an algebraic mistake in Ref.~\cite{KLP00}, but
the results of this paper are valid in the limit of negligible
magnetic shear considered here.}

\begin{align}
    \label{equ:disp}
    \nonumber
    \omega^{4} & -\left[k_\parallel^{2}(v_{A}^{2}+v_{SM}^{2})+
                   \omega_c^2\right]\omega^2\\[0.3\baselineskip]
               & +k_\parallel^{2}(k_\parallel^{2}-2\beta\kappa_\rho\kappa_c)
               v_{SM}^{2}v_{A}^{2}=0,
\end{align}

\noindent where $v_{SM}^2\simeq \beta v_A^2/(1+\beta)$ is the slow
magnetosonic square speed.

A necessary and sufficient condition of instability is

\begin{equation}\label{equ:crit}
  k_\parallel^{2}< 2\beta\kappa_\rho\kappa_c,
\end{equation}

\noindent which requires in particular $2\beta\kappa_\rho\kappa_c
>0$, a very-well known constraint for the existence of
pressure-driven instabilities. Note that Eq.~(\ref{equ:crit})
generalizes a criterion derived by Kadomtsev for the $m=1$ mode
\cite{KAD,Beg98,KLP00}. In the limit $k_\parallel\rightarrow 0$,
there is a small and a large root to the dispersion relation
Eq.~(\ref{equ:disp}), which, to leading order in $k_\parallel L_o$
are approximately given by

\begin{equation}
   \label{equ:omslow}
   \omega_{-}^2\simeq -v_{SM}^2\frac{2\beta\kappa_\rho\kappa_c v_A^2}{\omega_c^2}k_\parallel^2,
\end{equation}

\noindent and

\begin{equation}
    \label{equ:omalf}
    \omega_{+}^2\simeq \omega_c^2,
\end{equation}

\noindent Furthermore, in the limit $\kappa_\rho,
\kappa_c\rightarrow 0$, the large and small roots reduce to the
standard Alfv\'en and slow magnetosonic root, respectively. Only
one of these roots is unstable at a time, depending on the sign of
$\omega_c^2$.

Note that the large root $\omega_{+}^2$ satisfies the ordering
assumed in the derivation of the new criterion; furthermore, for
displacements $\xi_r$ satisfying the WKB approximation assumed in
the derivation of Eq.~(\ref{equ:disp}), Eq.~(\ref{omcrit}) gives
back Eq.~(\ref{equ:omalf}). Reversely, the large root mode cannot
satisfy the ordering needed to rederive Suydam's criterion, in
contrast to small root modes.

These arguments show that Suydam's criterion relates to the
stability of the slow magnetosonic branch, while the newly derived
criterion relates to the stability of the Alfv\'enic one, a
conclusion that can also be reached by an analysis of the global
modes behavior \cite{Long03}. Of the two branches, the Alfv\'enic
one the most dangerous, because it has the fastest growth rate, so
that the new criterion is probably the more relevant of the two to
ascertain the stability of MHD cylindrical equilibria.

The practical usefulness of this new criterion also follows from
the fact that $\omega^2\lesssim\omega_c^2/2$ always provides a
rather precise order of magnitude estimate of the unstable
pressure modes growth rates, even for low wavenumbers (down to
$m=1$, in fact \cite{LB03}); this feature also explains why the
most unstable pressure-driven modes have a growth rate nearly
independent of their wavenumber. Quite surprisingly, the new
criterion does not contain the magnetic shear. In fact, the
magnetic shear only role is to limit the mode width through
Eq.~(\ref{delta}) \cite{Long03}; it is because of this feature,
and not because of its influence on the growth rate, that the
magnetic shear can limit the disruptive power of Alfv\'enic
pressure-driven instabilities.

Under which conditions can one expect Eq.~(\ref{crit}) to be
satisfied ?

It is well-known that, in a Z-pinch equilibrium, $\omega_c^2
> 0$ is a constraint which is difficult to realize in practice. In
more realistic equilibria, however, one expects the longitudinal
component of the equilibrium magnetic field to decrease with
increasing radius, so that the gas pressure-gradient is relatively
weaker in such a configuration than in a Z-pinch equilibrium, a
feature which stabilizes the plasma, unless the pressure gradient
occurs only on a rather small radial scale. In fact, this is
precisely what happens in some of the most commonly considered
tokamaks or RFP configurations. Furthermore, fusion devices are
collisionless over the time-scales under which they usually
operate; the simplest (albeit crude) way to account for this is to
set the adiabatic index $\gamma=0$ in the problem, which
suppresses the compression stabilizing term. Therefore, either for
one or the other reason, pressure-driven interchange modes are
necessarily present in fusion devices' plasmas, down to low
azimuthal wave-numbers $m \gtrsim 1$ (in any case, large
wavenumber modes should be stabilized by finite Larmor radius
effects); furthermore, the growth rates are comparable to the
Alfv\'en time-scale once the plasma $\beta$ is of the order of a
significant fraction of unity. These features strongly suggest
that the role of pressure-driven interchange modes in the
generation of the sawtooth oscillations and edge localized modes
in tokamaks has been underestimated, a possibility which will be
more fully explored elsewhere. These modes might also play a role
in triggering the edge turbulent transport in fusion devices; on
the other hand, it is now well-known that a velocity shear can
both stabilize pressure-driven modes \cite{chiueh} and reduce the
MHD-driven edge transport \cite{TER}. A velocity shear can also
produce turbulent transport by itself through hydrodynamical
processes, but the flow curvature and the Coriolis force can
severely limit the induced turbulent transport
\cite{Long02,Long04}.

In what concerns astrophysical jets, most models studied in the
literature are cold, i.e., the equilibrium is not provided by the
balance between the pressure gradient and the magnetic tension due
to the azimuthal field, but by a balance between the inertial
force due to rotation and the magnetic tension
\cite{PP92,F97,CF00}. Destabilization by the rotation inertial
force is expected in this situation \cite{KLP00}, in spite of the
stabilizing action of the Coriolis force, with features very
similar to the ones described above. A more extensive study of
rotation-driven instabilities in astrophysical jets will be
presented elsewhere, while a comparison of usual pressure-driven
instabilities with the Kelvin-Helmholtz instability is performed
in Ref.~\cite{LB03}.

%\subsection
%\subsubsection{}

% If in two-column mode, this environment will change to single-column
% format so that long equations can be displayed. Use
%\begin{widetext}
%\end{widetext}


%\begin{acknowledgments}
%\end{acknowledgments}


%\bibliography{pyl}

\end{document}